\documentclass[12pt]{iopart}

\usepackage{amsfonts}
\usepackage{graphicx}
\eqnobysec

\usepackage{color}
\definecolor{darkgreen}{rgb}{0,0.6,0}

\newcommand{\misc}[1]{}

\begin{document}

\title{Non-ideal classical measurements and quantum measurements: a comparative study}

\author{Lars M. Johansen$^1$, Amir Kalev$^2$ and Pier A. Mello$^3$}
\address{$^1$Faculty of Technology, Buskerud University College,
N-3601 Kongsberg, Norway}

\address{$^2$Center for Quantum Information and Control, University of New Mexico, Albuquerque, NM 87131-0001, USA}

\address{$^3$Departamento de Sistemas Complejos, Instituto de
F\'{\i}sica, Universidad Nacional Aut\'{o}noma de M\'{e}xico,
M\'{e}xico, D.F. C.P. 04510}

\ead{lars.m.johansen@gmail.com, amirk@unm.edu, mello@fisica.unam.mx}

\begin{abstract}
Measurements on classical systems are usually idealized and assumed to have infinite precision. In practice, however, any measurement has a finite resolution.
We investigate the theory of non-ideal measurements in classical mechanics using a measurement probe with finite resolution. We use the von Neumann interaction model to represent the interaction between system and probe. We find that in reality classical systems are affected by measurement in a similar manner as quantum systems. 
In particular, we derive classical equivalents of L\"uders' rule, the ``collapse postulate", and the Lindblad equation. 
\end{abstract}


\maketitle

\section{Introduction}
\label{intro}

The process of measurement in quantum mechanics (QM) has been the cause of much controversy regarding its interpretation. In particular, the interpretation of the process of collapse of the wave-function in a measurement is highly controversial.

Measurements in classical mechanics (CM), on the other hand, are usually dealt with in the literature very briefly by assuming simply that they have infinite precision and do not affect the system; 
hence, the measurement process of classical systems is usually ignored (for a few notable exceptions, see Refs. \cite{wiseman,bartlett}). 
In practice, however, any measurement has a finite resolution. 
In this paper we study the effects of the finite resolution of a classical measurement by allowing the initial distribution of the measurement probe to be uncertain \cite{born,nauenberg},
i.e., a phase-space probability density distribution different from a Dirac delta function used for the infinite-precision-measurement case. 
This situation is quite generic, because the most general description of the probe is given by an extended phase-space probability density, particularly if the probe is a non-integrable system, or a part thereof \cite{prigogine}.
By using the von Neumann interaction model \cite{von_neumann} to represent the measurement process, we find that a non-ideal classical measurement 
may influence the system in a most profound manner, and
mimics a QM measurement in surprisingly detailed ways. We therefore attribute effects, such as L\"uders' rule, the collapse postulate, and the Lindblad equation, simply to the statistical nature of quantum theory, as their classical counterparts appear in a non-ideal, uncertain, classical measurement.

Although CM and QM have a formal similarity in their description of the dynamical evolution (in the sense that Poisson brackets of functions defined in phase-space appear in the former, 
while in the latter one finds commutators of operators defined in Hilbert space),
important differences may occur
because of the different spaces in which the two theories are defined.
For instance, the initial condition for the classical probability distribution in phase-space could be an arbitrary integrable non-negative function, while in QM there is no joint probability distribution of coordinate and momentum, and the two marginal distributions must fulfill the uncertainty principle. Thus, when carrying out the study of a classical measurement in terms of the von Neumann model mentioned above, one finds various relations which bear a similarity in structure to those occurring in QM, although, because of the above reasons, it is clear that one cannot, in general, expect a quantitative coincidence.
The purpose of the present paper is to uncover a number of such relations,
whose implications for the measurement process have not, to our knowledge, been pointed out in the past.
Some of these relations, describing the expectation value of observables and also the state of the system and the measurement probe, appear, at first sight, to be rather subtle and unexpected.

The comparison of various effects in classical and in quantum measurements is of great value, as we believe that it leads to a more profound understanding of the role played by CM and QM in the measurement process.

In Sec. \ref{sec:qm} we briefly review the theory of measurement in QM. By using the von Neumann measurement model (vNM), we derive L\"uders' rule, the Lindblad equation and a diffusion equation for the Wigner distribution. 
When going to the classical case in Sec. \ref{sec:cm}, we find a classical equivalent of these results, provided that the classical system is subject to measurement probes of finite resolution;
then it responds in a similar fashion as quantum systems.
We conclude in Sec. \ref{conclusions}, where we present, in Table~~\ref{tbl:sum}, a summary of the main results found in the paper that show a similar structure in CM and in QM.
We also include a number of appendices, where some of the derivations are presented in order not to lose the main flux of the paper. 
We remark that Sec. \ref{sec:cm} solves the CM problem in what might be called the ``Schr\"odinger picture";
when the quantity to be ``measured" is sufficiently simple, it turns out that one can gain an appreciable intuitive insight by solving the problem in the ``Heisenberg picture": this alternative derivation of some of the CM results is presented in~\ref{meas_CM_heisenberg}.

\section{Measurement in quantum mechanics}
\label{sec:qm}

In this section we re-derive some essential aspects of quantum measurements. This is done to serve as a point of reference for Sec. \ref{sec:cm}, where we derive the corresponding effects in classical-mechanical systems.  
The dynamical model of quantum measurements is given by  the vNM interaction Hamiltonian \cite{von_neumann} (see also Ref. \cite{peres_book} Ch. 12, Ref. \cite{ballentine} Ch. 9, and Ref. \cite{johansen-mello_pla08})
\begin{equation}
\hat{H} = \epsilon \delta (t-t_1) \hat{A} \hat{P} \;.
\label{H vNM QM}
\end{equation}
This Hamiltonian describes the  interaction between two quantum systems: 
the system proper $s$ and a probe $\pi$.
The probe and, in several applications, the system as well, are assumed, for simplicity, to be one-dimensional. The operator $\hat{A}$ represents the system observable of interest, $\hat{P}$ is the momentum of the probe, and $\epsilon$ represents the coupling strength.
Thus $\hat{H}$ introduces the first stage of the measurement process, or ``pre-measurement", explicitly in the QM description, through the system-probe coupling;
we shall speak, for short, of the {\em measurement of} $\hat{A}$. Through the system-probe entanglement we obtain information on the system observable $\hat{A}$;
it is some observable of the probe, not of the system proper, what is really detected by an experimental device: we shall then speak of the {\em detection} of the probe observable.
We assume that the evolution due to the free Hamiltonian may be neglected and that the interaction acts instantaneously at $t=t_1$. The density operator for the whole system, i.e.,  system proper and probe, for $t>t_1$ is given by
\begin{equation}
\hat{\rho}_{s\pi}'
= e^{-\frac{i}{\hbar}\epsilon \hat{A}\hat{P}} \hat{\rho}_s  \hat{\rho}_{\pi}  
e^{\frac{i}{\hbar}\epsilon \hat{A}\hat{P}} \; ,
\label{QM rho system+probe}
\end{equation}
where $\hat{\rho}_s$ and $\hat{\rho}_{\pi}$ are the density operators for the system and probe, respectively, at $t=0$. 
Throughout this paper we use primes to indicate quantities that are related to the evolved state, while unprimed quantities refer to the initial state.

\subsection{The marginal distribution of the probe position}
\label{pointer distribution}

After the interaction is over, we find \cite{johansen-mello_pla08}, for the marginal distribution of the probe position variable $Q$,
\begin{equation}
p'_{\pi}(Q) = \sum_n p_{s}(a_n) \; p_{\pi}(Q-\epsilon a_n),
\label{p(Q)QM}
\end{equation}
where $p_{s}(a_n)={\rm Tr}(\hat{\rho}_s \hat{\mathbb{P}}_{a_n})$ 
is the Born probability to find the eigenvalue $a_n$ of the observable $\hat{A}$
in the state $\hat{\rho}_s$
($\hat{\mathbb{P}}_{a_n}$ is the projector onto the system subspace associated with $a_n$),
and $p_{\pi}(Q-\epsilon a_n)
=\left\langle Q-\epsilon a_n|\hat{\rho}_{\pi} |Q-\epsilon a_n \right\rangle$. 
The observation of $\hat{Q}$, in order to extract the eigenvalues $a_n$ of the measured observable $\hat{A}$, permits discriminating the latter provided 
\begin{equation}
\frac{\sigma_Q}{
\epsilon} \ll | a_{n+1} - a_n |\;,
\label{discr_an}
\end{equation}
$\sigma_Q$ being the standard deviation of $Q$ in the state $\hat{\rho}_{\pi}$ of the probe.

From Eq. (\ref{p(Q)QM}) we find the average value of $Q$ after the interaction is over as
\begin{equation}
\frac{\langle \hat{Q} \rangle'}{\epsilon} 
= \langle \hat{A}  \rangle,
\label{<Q>QM}
\end{equation}
which is the expectation value of $\hat{A}$ before the interaction has been turned on.   
We have assumed that the original average probe position $\langle \hat{Q}  \rangle$ vanishes. 

As a particular example, if the observable is the position operator, i.e., 
$\hat{A}=\hat{x}$, the marginal distribution of the probe position $Q$ for $t>t_1$ is given by the convolution \cite{peres98}
\begin{equation}
p'_{\pi}(Q) = \int p_{s}(x) \; p_{\pi}(Q-\epsilon x) dx \; ,
\label{p(Q) QM x}
\end{equation}
where $p_{s}(x) = \langle x| \hat{\rho}_s |x\rangle$ is the original probability density for the result $x$ associated with the observable $\hat{x}$, and 
$p_{\pi}(Q-\epsilon x)=\left\langle Q-\epsilon x|\hat{\rho}_{\pi} |Q-\epsilon x \right\rangle$ 
is the original $Q$ probability density for the probe, displaced by the amount 
$\epsilon x$.
We may define the uncertainty in discriminating $x$'s as
\begin{equation}
\{{\rm uncertainty \; in \; discriminating} \; x^{,}{\rm s} \;
{\rm in \; the \; original \; state} \}
\sim \frac{\sigma_Q}{\epsilon} \; .
\label{QM-uncert-x}
\end{equation}

If the original state of the system is the convex combination
\begin{equation}
\hat{\rho}_s = p_{1}\hat{\rho}_{1,s} + p_2\hat{\rho}_{2,s},
\hspace{1cm} {\rm with} \hspace{1cm}  p_1 +  p_2 =1 \;,
\label{rhos=rhos1+rhos2 QM}
\end{equation}
$p_{s}(x)$, as well as the resulting $p'_{\pi}(Q)$ of 
Eq. (\ref{p(Q) QM x}), also split into two terms.
However, if the original state of the system is the pure state
\begin{equation}
|\psi\rangle_s = \alpha |\psi_1\rangle_s + \beta |\psi_{2}\rangle_s \; ,
\label{psis=psi1s+psi2s}
\end{equation}
the probability density at $t=0$, $p_s(x)$,
\begin{equation}
p_s(x)
= |\alpha|^2 |\psi_{1,s}(x)|^2   + |\beta|^2 |\psi_{2,s}^{(2)}|^2
+ \left\{\alpha \beta^{*} \psi_{1,s}(x)\left[\psi_{2,s}(x)\right]^{*} 
+ c.c. \right\}
\label{W(x) with interference}
\end{equation}
exhibits interference terms, and so does the resulting $p'_{\pi}(Q)$ of 
Eq. (\ref{p(Q) QM x}).

\subsection{The reduced density operator of the system proper}
\label{reduced rho-s QM}

The reduced density operator of the system proper $s$ after its interaction with the probe is given by
\begin{eqnarray}
\hat{\rho}_s' &=& {\rm Tr}_{\pi} \hat{\rho}_{s\pi}' \nonumber \\ 
&=& \int dP \langle P \mid e^{-\frac{i}{\hbar} \epsilon \hat{A} \hat{P}} \hat{\rho}_{\pi} \hat{\rho}_s 
e^{\frac{i}{\hbar} \epsilon \hat{A} \hat{P}} \mid P \rangle \nonumber \\ 
&=& \int dP \langle P | \hat{\rho}_{\pi} | P \rangle 
e^{-\frac{i}{\hbar} \epsilon \hat{A} P}  \hat{\rho}_s 
e^{\frac{i}{\hbar} \epsilon \hat{A} P}.
\label{eq:reduced1}
\end{eqnarray}
To this end, we shall assume that the probe has a zero-centered Gaussian momentum distribution
\begin{equation}
   \langle P | \hat{\rho}_{\pi} | P \rangle = \frac{1}{\sqrt{2 \pi} \, \sigma_P}
 e^{-\frac{P^2}{2 \sigma_P^2}}.
 \label{eq:probe_momentum}
\end{equation}

\subsubsection{The Lindblad equation}

We first recall the Baker-Hausdorff lemma 
(see, e.g., Ref. \cite{messiah}, p. 339)
\begin{equation}
 e^{\xi \hat{A}} \hat{B} e^{-\xi \hat{A}} = \hat{B} + \xi [\hat{A},\hat{B}] + \frac{\xi^2}{2!} [\hat{A},[\hat{A},\hat{B}]] + ... =  \left [ \sum_{n=0}^\infty \frac{(\xi \hat{\cal{A}})^n}{n!} \right ] \hat{B},
\end{equation}
where the action of the ``superoperator" $\hat{\cal{A}}$ is defined as
\begin{equation}
   \hat{\cal{A}} \hat{B} = [\hat{A},\hat{B}].
\end{equation}
Thereby, we may rewrite Eq. (\ref{eq:reduced1}) as
\begin{equation}\label{eq:reduced2}
\hat{\rho}_s' 
= \left [ \sum_{n=0}^\infty  \langle\hat{P}^n\rangle_\pi 
\frac{(-\frac{i}{\hbar} \epsilon \hat{\cal{A}})^n}{n!} \right ] \hat{\rho}_s
= \left\langle e^{-\frac{i}{\hbar} \epsilon P \hat{\cal{A}}} \right\rangle_{\pi} \hat{\rho}_s 
\; ,
\end{equation}
where we formally wrote the result as an exponential, the average being performed over the probe-momentum distribution, whose moments are
\begin{equation}
   \langle\hat{P}^n\rangle_\pi = \int dP \, P^n \, \langle P | \hat{\rho}_{\pi} | P \rangle.
\end{equation}
The result (\ref{eq:reduced2}) is valid for an arbitrary probe-momentum distribution.
For the Gaussian case of Eq. (\ref{eq:probe_momentum}) we find
\begin{equation}
\langle\hat{P}^n\rangle_\pi = \left \{  
\begin{array}{ccc}
\frac{n! \, \sigma_P^n}{\left ( \frac{n}{2} \right )! \, { 2^{n/2}}} & \textrm{for} & n=0,2,4, ... \\ 0 & \textrm{for} & n=1,3,5, ...
\end{array}
\right .
\end{equation}
We insert this into Eq. (\ref{eq:reduced2}), and obtain
\begin{eqnarray}
\hat{\rho}_s' 
= \sum_{m=0}^\infty  \frac{1}{m!}\left (- \frac{\tau}{\hbar^2} \hat{\cal{A}}^2 \right )^m\hat{\rho}_s 
= e^{- \frac{\tau}{\hbar^2} \hat{\cal{A}}^2} \hat{\rho}_s,
\label{QM rho's 1}
\end{eqnarray}
where
\begin{equation}
 \tau=\frac{1}{2}\left(\epsilon \sigma_P\right)^2.
\label{tau}
\end{equation}
Either when $\sigma_P =0$, or when the observable $\hat{A}$ and $\hat{\rho}_s$ commute, 
$[\hat{A}, \hat{\rho}_s]=0$,
the reduced state of the system does not change as a result of 
its interaction with the probe, i.e., $\hat{\rho}'_s = \hat{\rho}_s$.

By differentiating Eq. (\ref{QM rho's 1}) with respect to $\tau$, we obtain the Lindblad equation
\cite{lindblad,gorini}
\begin{equation}
 \frac{\partial \hat{\rho}_s'}{\partial \tau} 
= - \frac{1}{\hbar^2} \hat{\cal{A}}^2 \hat{\rho}_s' 
= - \frac{1}{\hbar^2} [\hat{A},[\hat{A},\hat{\rho}_s']].
\label{lindblad}
\end{equation}
The Lindblad equation is widely used to describe the non-unitary time evolution of open quantum systems. 
It ensures that the evolution is completely positive
(i.e., the density operator remains positive)
and trace preserving.

Results (\ref{QM rho's 1}) and (\ref{lindblad}) are valid for a spectrum of $\hat{A}$ which could be either discrete or continuous.

\subsubsection{L\"uders rule}

i) We now assume that the observable $\hat{A}$ has the discrete spectral resolution
\begin{equation}
	\hat{A} = \sum_n a_n \hat{\mathbb{P}}_{a_n} \; ,
\label{spectral-repr-A}
\end{equation}
where the projectors satisfy
\begin{equation}
\hat{\mathbb{P}}_{a_m} \hat{\mathbb{P}}_{a_n} 
= \delta_{mn} \hat{\mathbb{P}}_{a_n},\;\; \sum_m \hat{\mathbb{P}}_{a_n} = \hat{1}.
\end{equation}
This enables us to write the reduced state (\ref{eq:reduced1}) as
\cite{{johansen-mello_pla08}}
\begin{equation}
\hat{\rho}_s' = \sum_{mn} g_{mn} \hat{\mathbb{P}}_{a_m} \hat{\rho}_s 
\hat{\mathbb{P}}_{a_n} \; ,
\label{rho's}
\end{equation}
where
\begin{equation}
	g_{mn} = \int dP \langle P | \hat{\rho}_{\pi} | P \rangle 
	e^{-i \frac{\epsilon}{\hbar} (a_m - a_n) P}.
\end{equation}
From Eq. (\ref{rho's}) we see that
\begin{equation}
{\rm Tr}_s (\hat{\rho}'_{s} \mathbb{P}_{a_{n}})
= {\rm Tr}_s (\hat{\rho}_{s} \mathbb{P}_{a_{n}}) \;,
\label{w_a_nu unchanged}
\end{equation}
meaning that the probability distribution of the eigenvalues $a_n$ of $\hat{A}$ is not altered by the measurement of $\hat{A}$:
the reason for this is that the Hamiltonian of Eq. (\ref{H vNM QM}) commutes with the operator $\hat{A}$. 

By using the Gaussian distribution of Eq. (\ref{eq:probe_momentum}), $g_{mn}$ reduces to 
\begin{equation}
	g_{mn} = e^{- \frac{\tau}{\hbar^2} (a_m-a_n)^2}.
\end{equation}
It is readily seen that, in the limit of very large coupling, 
\begin{equation}
	\lim_{\tau \rightarrow \infty} g_{mn} = \delta_{mn},
\end{equation}
whereby
\begin{equation}
	\lim_{\tau \rightarrow \infty} \hat{\rho}_s' = \sum_{n} 
\hat{\mathbb{P}}_{a_n} \hat{\rho}_s \hat{\mathbb{P}}_{a_n}.
\label{QM lueders}
\end{equation}
This is {\it L\"uders rule} \cite{lueders} for a {\it non-selective} projective measurement \cite{lars07}.

Eq. (\ref{QM lueders}) can be written, in the absence of degeneracy, as
\begin{equation}
\lim_{\tau \to \infty} \hat{\rho}_{s}'
= \sum_{n} |a_n \rangle p_{s}(a_n)\langle a_n | \; .
\label{QM lueders 1}
\end{equation}
Thus, as a result of the measurement of the observable $\hat{A}$, the reduced density operator for the system proper becomes, in the strong coupling limit, 
{\it a function of the operator $\hat{A}$} (and of $\hat{\rho}_s$), 
the RHS Eq. (\ref{QM lueders 1}) being its spectral representation.
An illustrative example of this behavior is discussed in  
\ref{example of final system reduced rho QM}.

ii) We now assume that the observable $\hat{A}$ is the position operator $\hat{x}$, and thereby has a continuous spectrum. 
Then the reduced density operator for the system after its interaction with the probe can be written as
\begin{equation}
\hat{\rho}'_{s}
= \int \int g^{(\hat{x})}_{x x'} \;
\mathbb{P}_{x} \hat{\rho}_s \mathbb{P}_{x'} 
dx dx' \; ,
\label{QM_Lueder x}
\end{equation}
where now
\begin{equation}
\mathbb{P}_{x} = |x \rangle \langle x | \ ,
\label{P_x}
\end{equation}
and
\begin{equation}
g_{x x'}
=e^{-\frac{\tau}{\hbar^2} (x - x')^2} \; ,
\label{g_xx'}
\end{equation}
$\tau$ being defined in Eq. (\ref{tau}).
We may take, as a measure of the {\it disturbance on the system produced by the 
measurement of $\hat{x}$}, the quantity
\begin{equation}
\{ {\rm disturbance \; on \; the \; system} \}
\sim 
\frac{\epsilon}{\hbar} \sigma_P \sqrt{{\rm var}(\hat{x})}
\label{QM disturbance on system}
\end{equation}
where ${\rm var}(\hat{x})$ is the variance of $\hat{x}$ in the initial state of the system.
Notice again (see comment right after Eq. (\ref{tau})) that the disturbance caused on the system proper vanishes if $\sigma_P=0$.

\subsubsection{The Wigner transform of the reduced density operator for the system}
\label{Wigner_rho_s}

For future comparisons with the classical case, it is useful to analyze
the Wigner transform (WT) of the density operator $\hat{\rho}$, defined as
\cite{schleich}
\begin{equation}
W_{\hat{\rho}}(q,p)
=
\int_{-\infty}^{\infty} e^{-\frac{i}{\hbar}py}\Bigl\langle q + \frac{y}{2} \Bigl | \hat{\rho} \Bigr | q - \frac{y}{2} \Bigr\rangle
dy,
\label{W_rho}
\end{equation}
which is normalized according to the rule
\begin{equation}
\int_{-\infty}^{\infty} \int_{-\infty}^{\infty}
W_{\hat{\rho}}(q,p)\frac{dq dp}{2 \pi \hbar} = 1 \; .
\label{int_W_rho}
\end{equation}

In the particular case of the measurement of $A(\hat{x})$, the WT of the reduced density operator for the system proper after the interaction is over, 
$\hat{\rho}'_s$ of Eq. (\ref{QM_Lueder x})
generalized to $A(\hat{x})$, is found to be
\begin{equation}
W_{\hat{\rho}'_{s}}(q,p)
= 
\int_{-\infty}^{\infty} 
e^{- \frac{\tau}{\hbar^2}\left[A(q+\frac{y}{2})-A(q-\frac{y}{2})\right]^2} 
e^{-\frac{i}{\hbar}py}
\Bigl\langle q + \frac{y}{2} \Bigl | 
\hat{\rho}_s \Bigr |q-\frac{y}{2} \Bigr\rangle 
\; dy ,
\label{WT rho(t>t1)}
\end{equation}
in terms of the density operator $\hat{\rho}_s$ of the system proper prior to the interaction with the probe.
From Eq. (\ref{WT rho(t>t1)}) we see that $W_{\hat{\rho}'_{s}}(q,p)$ satisfies the equation
\begin{equation}
\frac{\partial W_{\hat{\rho}'_{s}}(q,p)}{\partial \tau}
=  \left[\frac{1}{i\hbar}\Delta A\left(q, i\hbar\frac{\partial}{\partial p}\right)\right]^2
W_{\hat{\rho}'_{s}}(q,p) \; ,
\label{diff_eqn_WT}
\end{equation}
where we have defined
\begin{equation}
\Delta A(q,y) 
= A\left(q+\frac{y}{2}\right)-A\left(q-\frac{y}{2}\right) \; ,
\label{DeltaA}
\end{equation}
which is clearly an odd function of $y$, and its square an even function of $y$.
Thus the RHS of Eq. (\ref{diff_eqn_WT}) contains even-order derivatives in the variable $p$.
 
It is interesting to examine the solution (\ref{WT rho(t>t1)}) for large coupling $\tau$.
As $\tau$ increases, the main contribution to the WT (\ref{WT rho(t>t1)}) comes from smaller $y${'}s (because then the square bracket in the exponential becomes small), so that in order to see a $p$ dependence we need to go to larger $p$'s. 
Thus the resulting $W_{\hat{\rho}'_s}(q,p)$ {\it looks ever more like a function of $q$ only}. 
This behavior is consistent with that found in relation with 
Eq. (\ref{QM lueders 1}).

\subsection{Selective projective measurements}
\label{sel-proj-meas}
Using the spectral representation (\ref{spectral-repr-A}) for the observable $\hat{A}$, we can write the density operator of the system plus probe after the interaction is over, Eq. (\ref{QM rho system+probe}), as
\begin{equation}
\hat{\rho}_{s\pi}'
= \sum_{n,n'}\mathbb{P}_{a_n} \hat{\rho}_s \mathbb{P}_{a_{n'}}
\left(
e^{-\frac{i}{\hbar}\epsilon a_n \hat{P}}   \hat{\rho}_{\pi}  
e^{\frac{i}{\hbar}\epsilon a_{n'}\hat{P}} 
\right).
\label{QM rho system+probe 1}
\end{equation}
The expectation value of an arbitrary observable $\hat{O}_s$ of the system proper, conditioned on the measured value $Q$ of the probe position, corresponds to the selection of the subensemble for a fixed $Q$ and is given by 
[$p'_{\pi}(Q)$ being given in Eq. (\ref{p(Q)QM})]
\numparts
\begin{eqnarray}
E'(\hat{O}_s | Q)
&=& \frac{\sum_{n,n'} 
{\rm Tr}_s \left(\mathbb{P}_{a_n} \hat{\rho}_s \mathbb{P}_{a_{n'}} \hat{O}_s \right)
{\rm Tr}_{\pi} \left(
e^{-\frac{i}{\hbar}\epsilon a_n \hat{P}}   \hat{\rho}_{\pi}  
e^{\frac{i}{\hbar}\epsilon a_{n'}\hat{P}} 
\mathbb{P}_Q \right)} {p'_{\pi}(Q)} \; ,
\label{E(O|Q) 1 a}  \\
&=& \frac{\sum_{n,n'} 
{\rm Tr}_s \left(\mathbb{P}_{a_n} \hat{\rho}_s \mathbb{P}_{a_{n'}} \hat{O}_s \right)
\left\langle Q -\epsilon a_n 
\big|
\hat{\rho}_{\pi}  
\big| Q  -\epsilon a_{n'} \right\rangle} 
{\sum_n {\rm Tr}_s(\hat{\rho}_s \hat{\mathbb{P}}_{a_n}) \; p_{\pi}(Q-\epsilon a_n)} \; ,
\label{E(O|Q) 1 b}   \\
&=&{\rm Tr}_s \left(\hat{\rho}'_s\big|_Q \;  \hat{O}_s \right) \; .
\end{eqnarray}
\label{E(O|Q) 1}
\endnumparts
We have defined the {\it conditional reduced density operator for the system}:
\begin{equation}
\hat{\rho}'_s\big|_Q
\equiv
\frac{\sum_{n,n'} 
\mathbb{P}_{a_n} \hat{\rho}_s \mathbb{P}_{a_{n'}} 
\left\langle Q -\epsilon a_n 
\big|
\hat{\rho}_{\pi} 
\big| Q  -\epsilon a_{n'} \right\rangle} 
{\sum_n {\rm Tr}_s(\hat{\rho}_s \hat{\mathbb{P}}_{a_n}) \; p_{\pi}(Q-\epsilon a_n)} \; .
\label{rho's|Q}
\end{equation}
Assume now that the probe is described by the pure state 
$\hat{\rho}_{\pi} =|\chi \rangle \langle \chi|$.
We also assume the coupling to be so strong that 
$p_{\pi}(Q)=|\chi(Q-\epsilon a_n)|^2$ is very narrow,
$\sigma_Q \ll \epsilon|a_n - a_{n'}|$, so that adjacent distributions, for $a_n$ and $a_{n+1}$, do not ovelap.
Then
\numparts
\begin{eqnarray}
\left\langle Q -\epsilon a_n 
\big|
\hat{\rho}_{\pi}  
\big| Q  -\epsilon a_{n'} \right\rangle
&=& \chi(Q -\epsilon a_n ) \chi^{*}(Q -\epsilon a_{n'}) \\
&\approx& |\chi(Q-\epsilon a_n)|^2 \delta _{a_n a_{n'}} \\
&=&p_{\pi}(Q-\epsilon a_n) \delta _{a_n a_{n'}} ,
\end{eqnarray}
\endnumparts
and (\ref{rho's|Q}) reduces to
\begin{equation}
\hat{\rho}'_s\big|_Q
\approx
\frac{\sum_{n} 
\mathbb{P}_{a_n} \hat{\rho}_s \mathbb{P}_{a_{n}} \;
p_{\pi}(Q-\epsilon a_n)}
{\sum_n {\rm Tr}_s(\hat{\rho}_s \hat{\mathbb{P}}_{a_n}) \; p_{\pi}(Q-\epsilon a_n)} \; .
\label{rho's|Q 1}
\end{equation}
Since the probe position $Q$ is at our disposal, suppose we choose
\begin{equation}
Q= \epsilon a_{\nu}.
\label{particular Q}
\end{equation}
Then (\ref{rho's|Q 1}) becomes
\begin{equation}
\hat{\rho}'_s\big|_{Q=\epsilon a_{\nu}}
\approx
\frac{\mathbb{P}_{a_{\nu}} \hat{\rho}_s \mathbb{P}_{a_{\nu}}}
{Tr_s(\hat{\rho}_s \hat{\mathbb{P}}_{a_{\nu}})} \; .
\label{rho's|Q 2}
\end{equation}
This is {\it L\"uders rule} \cite{lueders} for a {\it selective} projective measurement \cite{lars07},
sometimes referred to as ``collapse" of the state.

\section{Measurement in classical mechanics}
\label{sec:cm}

We consider a classical system, endowed, for simplicity, with only one degree of freedom.
Its state at $t<t_1$ is defined by the phase-space density $\rho_s(q,p)$, $q$ and $p$ being the corresponding canonical position and momentum.
We wish to obtain information about the system by coupling it to a probe and detecting some property of the latter. The probe is also endowed with one degree of freedom and is described, for $t<t_1$, by the phase-space density $\rho_{\pi}(Q,P)$, $Q$ and $P$ being the probe canonical position and momentum.

In order to carry on this program, we consider a system-probe interaction as the CM counterpart of the QM vNM of Eq. (\ref{H vNM QM}), 
assuming that we wish to measure the system observable $A(q,p)$ 
(Ref. \cite{peres_book}, Ch. 12), i.e., 
\begin{equation}
H(t) = \epsilon \delta (t-t_1)A(q,p) P,
\label{V_vNM CM}
\end{equation}
again neglecting the intrinsic evolution of the system and the probe.

The evolution of the state of the total system --system proper plus probe-- is governed by Liouville's equation
\begin{equation}\label{Liouv_eqn}
\frac{\partial \rho_{s \pi}(t)}{\partial t} = \left[H, \rho_{s \pi}(t)\right]_{PB} \equiv \hat{H}_{op} \rho_{s \pi}(t) \; .
\end{equation}
Here, $\rho_{s \pi}(t) \equiv \rho_{s \pi}(q,p,Q,P;t)$ is the phase-space density of the total system at time $t$, and $\left[A,B\right]_{PB}$ denotes the Poisson bracket of $A$ and $B$.
We have also defined the operator 
[the index $i$ runs over the coordinates (or momenta) of the system and  the probe]
\numparts
\begin{eqnarray}
\hat{H}_{op} &=& \sum_i \left[ \frac{\partial H}{\partial q_i} \frac{\partial}{\partial p_i}
-\frac{\partial H}{\partial p_i} \frac{\partial}{\partial q_i} \right]
\\
&=& \epsilon \delta (t-t_1)
\left[
\left(  \frac{\partial A}{\partial q} \frac{\partial}{\partial p} 
- \frac{\partial A}{\partial p} \frac{\partial}{\partial q} \right) P
-A(q,p)\frac{\partial}{\partial Q}  
\right] \\
&\equiv& \epsilon \delta (t-t_1) \hat{{\cal K}} \; ,
\end{eqnarray}
\label{Hop}
\endnumparts
related to the so-called Liouville operator $\hat{L}$ by $\hat{L}=i\hat{H}_{op}$
\cite{reichl}.
We have further defined the operator
\begin{equation}\label{K}
\hat{{\cal K}} 
=\left(  \frac{\partial A}{\partial q} \frac{\partial}{\partial p} 
- \frac{\partial A}{\partial p} \frac{\partial}{\partial q} \right) P
-A(q,p)\frac{\partial}{\partial Q} \equiv \hat{A}_{op}P - A(q,p)\frac{\partial}{\partial Q} \; ,
\end{equation}
with 
\begin{equation}
\hat{A}_{op}
=\frac{\partial A}{\partial q} \frac{\partial}{\partial p}
-\frac{\partial A}{\partial p} \frac{\partial}{\partial q} \; .
\label{Aop}
\end{equation}
Assuming for the total state the initial condition 
$\rho_{s\pi}(q,p,Q,P)= \rho_s(q,p) \rho_{\pi}(Q,P)$, the state at time $t$ is given by
\begin{equation}
\rho_{s \pi}(q,p,Q,P;t) = 
e^{\epsilon \; \theta(t-t_1) \hat{{\cal K}}} \rho_s(q,p) \rho_{\pi}(Q,P) \; ,
\label{rho_t}
\end{equation}
so that, after the interaction is over, i.e., for $t>t_1$, the new state is 
\begin{equation}
\rho'_{s \pi}(q,p,Q,P) = 
e^{\epsilon \; \hat{{\cal K}}} \rho_s(q,p) \rho_{\pi}(Q,P) \; .
\label{rho_t>t1}
\end{equation}
It is shown in  \ref{proof_[AP,AdQ]} that the two operators 
$\hat{A}_{op}P$ and
$A(q,p)\partial / \partial Q$ occurring in Eq. (\ref{K}) commute, i.e.,
\begin{equation}
\left[ \hat{A}_{op}P, \; A(q,p)\frac{\partial}{\partial Q} \right] =0\; ,
\label{[AP,AdQ]}
\end{equation}
implying that one can write the evolution operator in Eq. (\ref{rho_t>t1}) in the two following equivalent ways
\numparts
\begin{eqnarray}
e^{\epsilon \; \hat{{\cal K}}}
&=& e^{\epsilon \hat{A}_{op}P} 
e^{-\epsilon A(q,p)\frac{\partial}{\partial Q}}
\label{expK a}
\\
&=& e^{-\epsilon A(q,p)\frac{\partial}{\partial Q}} \;
e^{\epsilon \hat{A}_{op}P} \; .
\label{expK b}
\end{eqnarray}
\endnumparts
Using the form (\ref{expK a}), the state of Eq. (\ref{rho_t>t1}) can be expressed as
\numparts
\begin{eqnarray}
\rho'_{s \pi}(q,p,Q,P) 
= e^{\epsilon \hat{A}_{op}(q,p)P}
\Big[ 
\rho_s(q,p) \; \rho_{\pi}(Q - \epsilon A(q,p),P) 
\Big] \; ,
\label{rhot>t1 a}
\end{eqnarray}
while using the form (\ref{expK b}), the same state can be written as
\begin{eqnarray}
\rho'_{s \pi}(q,p,Q,P) 
= \Big[ e^{\epsilon \hat{A}_{op}(q,p)P}  \rho_s(q,p) \Big]
\rho_{\pi}(Q - \epsilon A(q,p),P) \; .
\label{rhot>t1 b}
\end{eqnarray}
\label{CM rho'}
\endnumparts
This is the CM equivalent of the QM result of Eq. (\ref{QM rho system+probe}).

\subsection{The marginal distribution of the probe}
\label{p(Q) single meas}

From Eq. (\ref{rhot>t1 a}) we can write, for the marginal distribution of the probe after the interaction with 
the system
\begin{equation}
\rho'_{\pi}(Q,P) 
= \int \int e^{\epsilon \hat{A}_{op}(q,p)P}
\Big[ 
\rho_s(q,p) \; \rho_{\pi}(Q - \epsilon A(q,p),P) 
\Big] dq dp.
\label{rho(Q,P) 1}
\end{equation}
This expression has the form
\begin{equation}
\rho'_{\pi}(Q,P) 
= \int \int e^{\epsilon \hat{A}_{op}(q,p)P}
F(q,p;Q,P) dq dp ,
\label{rho(Q,P) 2}
\end{equation}
where $F(q,p;Q,P)$ has the property of vanishing as 
$|q| \to \infty$, or $|p| \to \infty$.
If we expand the exponential in (\ref{rho(Q,P) 2}), the second term in the expansion, i.e., the term of order $\epsilon$,
is proportional to
\begin{eqnarray}
\int \int \hat{A}_{op}(q,p)
F(q,p;Q,P) dq dp 
&=& \int \int 
\left(\frac{\partial A}{\partial q} \frac{\partial F}{\partial p}
- \frac{\partial A}{\partial p} \frac{\partial F}{\partial q} \right)
dq dp 
\nonumber \\   
&=& \int \int 
\left( -\frac{\partial^2 A}{\partial p \partial q} 
+ \frac{\partial^2 A}{\partial q \partial p} \right)
F(q,p;Q,P) dq dp 
\nonumber \\
&=& 0,
\label{rho(Q,P) 3}
\end{eqnarray}
where we performed an integration by parts.
Similarly, we find that higher-order terms vanish, so that only the first term in the expansion of the exponential survives, with the result
\begin{equation}
\rho'_{\pi}(Q,P)
= \int \int \rho_s(q,p) \; \rho_{\pi}(Q - \epsilon A(q,p),P) dq dp.
\label{rhoM}
\end{equation}

We now use the final probability density of the probe to extract information on the system proper.
From the expression (\ref{rhoM}) we find the marginal distribution of the probe position $Q$ after the interaction has acted ($t>t_1$) as
\begin{equation}
\rho'_{\pi}(Q)
= \int \int \rho_s(q,p) \; \rho_{\pi}(Q - \epsilon A(q,p)) dq dp \; .
\label{rho'(Q)}
\end{equation}

The average of the probe position $Q$ after the interaction is over, assuming that it vanishes at $t=0$, is then given by
\begin{equation}
\frac{\langle Q \rangle'}{\epsilon} 
= \langle A  \rangle,
\label{<Q>'CM}
\end{equation}
which is the expectation value of $\hat{A}$ before the interaction has been turned on.   

We notice the similarity {\em in structure} of the CM Eq. (\ref{rho'(Q)}) and the QM one, Eq. (\ref{p(Q)QM}).
The results (\ref{<Q>'CM}) and (\ref{<Q>QM}) are actually identical.

A case of particular interest is the measurement of the system position $q$. Then $A(q,p)=q$
and Eq. (\ref{rho'(Q)}) reduces to the convolution \cite{peres98}
\begin{equation}
\rho'_{\pi}(Q)
= \int \rho_s(q) \; \rho_{\pi}(Q - \epsilon q) dq ,
\label{rho(Q) A=q}
\end{equation}
where $\rho_s(q)$ is the marginal position distribution for the system.
This result should be compared with the QM one, Eq. (\ref{p(Q) QM x}).
The convolution (\ref{rho(Q) A=q}) does not allow resolving details of $\rho_s(q)$ finer than 
$\sigma_Q / \epsilon$, $\sigma_Q$ being the width of the distribution $p_{\pi}(Q)$ at $t=0$. 
This feature holds both in QM and in CM.
As an example, assume that the $q$-marginal of the original system state is given by
\begin{equation}
\rho_s(q) = \frac12 [\delta (q-q_0) + \delta (q-q_1)].
\label{rho_s=d(q-q0)}
\end{equation}
Then the resulting distribution of the probe position is, from 
Eq. (\ref{rho(Q) A=q})
\begin{equation}
\rho'_{\pi}(Q)
= 
\frac12 [\rho_{\pi}(Q - \epsilon q_0) + \rho_{\pi}(Q - \epsilon q_1)].
\label{rho(Q) A=q 1}
\end{equation}
The original probe-position distribution $\rho_{\pi}(Q)$ has been broken up into two distributions 
centered at $\epsilon q_0$ and $\epsilon q_1$, each with a width $\sigma_Q$.
We see that $\rho'_{\pi}(Q)$ permits discriminating the two values $q_0$, $q_1$, provided 
$\sigma_Q / \epsilon \ll |q_1-q_0|$.

We write, schematically,
\begin{equation}
\{{\rm uncertainty \; in \; discriminating} \; q^{,}{\rm s}
{\rm \; in \; the \; original \; distribution} \}
\sim \frac{\sigma_Q}{\epsilon} , \;\;
\label{CM-uncert-q}
\end{equation}
the corresponding QM result being given in Eq. (\ref{QM-uncert-x}).
We conclude that {\it a dispersion $\sigma_Q$ in the original probe position produces, both in CM and in QM, an uncertainty in discriminating the values of the observable to be measured}.

If the original state of the system is split as
\begin{equation}
\rho_s(q,p) = p_1 \rho_{1,s}(q,p) + p_2 \rho_{2,s}(q,p) \; ,
\label{rhos=rho1s_rho2s}
\end{equation}
the situation is like the QM one, Eq. (\ref{rhos=rhos1+rhos2 QM}).
There is, however, an obvious difference with the QM results (\ref{psis=psi1s+psi2s}) and (\ref{W(x) with interference}), as there is no equivalent of a ``probability amplitude" in CM, the result being the absence of interference terms.

\subsection{The marginal distribution of the system proper}
\label{rho'(q,p)}

To obtain the marginal distribution of the system proper after the interaction has acted ($t>t_1$), we integrate Eq. (\ref{rhot>t1 b}) over $Q$ and $P$ and obtain
\begin{eqnarray}
\rho'_{s}(q,p) 
&=& \int dP \Big[ e^{\epsilon \hat{A}_{op}(q,p)P} \rho_s(q,p) \Big]
\int dQ \; \rho_{\pi }(Q - \epsilon A(q,p),P)
 \nonumber\\
&=& \int dP \Big[ e^{\epsilon \hat{A}_{op}(q,p)P} \rho_s(q,p) \Big] \rho_{\pi }(P)
\nonumber\\
&=&\Big[ \int dP e^{\epsilon \hat{A}_{op}(q,p)P} \rho_{\pi }(P) \Big]\rho_s(q,p),
\label{rho_s_t>t1 1}
\end{eqnarray}
where $\rho_{\pi }(P)$ is the original marginal distribution of the probe momentum.
For $\rho_{\pi }(P)$ we shall assume a zero-centered Gaussian distribution with  variance $\sigma_P^2$, i.e.,
\begin{equation}
\rho_{\pi}(P)
=\frac{e^{-\frac{P^2}{2\sigma_P^2}}}{\sqrt{2 \pi \sigma_P^2 }}.
\label{rho(P)}
\end{equation}
The integral on the last line in Eq. (\ref{rho_s_t>t1 1}) can be evaluated by expanding the exponential, with the result
\begin{equation}
\int dP e^{\epsilon \hat{A}_{op}(q,p)P} 
\frac{e^{-\frac{P^2}{2\sigma_P^2}}}{\sqrt{2 \pi \sigma_P^2 }}
=e^{\frac12 \epsilon^{2}\sigma_P^2  \hat{A}_{op}^{2}} ,
\label{rho'_s 0}   
\end{equation}
where we have used the identity $\frac{(2n)!}{(2n-1)!!}= n! 2^n$.
We thus find, for the marginal distribution of the system proper after the interaction has ceased to act,
\begin{equation}
\rho'_{s}(q,p)
=  e^{\tau \hat{A}_{op}^2} \; \rho_s(q,p) ,
\label{rho'_s 1}
\end{equation}
where $\tau$ is defined in Eq. (\ref{tau}).
Notice that the classical result just obtained, Eq. (\ref{rho'_s 1}), has a structure similar to the QM one given in Eq. (\ref{QM rho's 1}).
Just as in the QM case [see comment right after Eq. (\ref{tau})],
the marginal state of the system is not changed by its interaction with the probe, i.e., $\hat{\rho}'_s(q,p) = \hat{\rho}_s(q,p)$, either when 
$\sigma_P =0$, or when $\left[A(q,p), \rho_s(q,p)\right]_{PB}=0$.

We thus see that {\it a dispersion $\sigma_P$ in the original probe momentum generates, both in CM and in QM [see Eqs. (\ref{QM rho's 1}) and (\ref{QM disturbance on system})], a disturbance in the state of the system proper}. 

As discussed in the Introduction, in a more general situation we would have a dispersion in both probe position $Q$ and momentum $P$. 
If, in the above analysis, instead of $Q$ and $P$ we had the pair of canonically conjugate variables $\bar{Q}(Q,P)$ and $\bar{P}(Q,P)$, a dispersion in $Q$ and $P$ would cause i) a dispersion in $\bar{P}$ which, in turn, would cause a disturbance in the state of the system proper, and
ii) a dispersion in $\bar{Q}$ which, according to the previous subsection, would cause an uncertainty in discriminating the values of the observable to be measured.

Differentiating Eq. (\ref{rho'_s 1}) with respect to $\tau$, we obtain
\numparts
\begin{eqnarray}
\frac{\partial \rho'_{s} (q,p)}{\partial \tau}
&=& \hat{A}_{op}^2 \rho'_{s}(q,p) \; ,  
\label{CM diffus_eqn a}  \\
&=& \Big[A(q,p), \big[A(q,p), \rho'_{s}(q,p)\big]_{PB}\Big]_{PB} \; ,
\label{CM diffus_eqn b}
\end{eqnarray}
\label{CM diffus_eqn}
\endnumparts
a diffusion equation which is the 
{\it classical counterpart of the QM Lindblad equation} (\ref{lindblad}).
It can be obtained from the latter equation by replacing the commutators by Poisson brackets according to the standard rule
\begin{equation}
[\hat{F}, \hat{G}]  \Rightarrow i \hbar [F,G]_{PB}           \; .
\label{PB->comm}
\end{equation}
We now study the ``evolution" with $\tau$ of $\rho'_{s\pi}(q,p)$ in some simple, yet illuminating cases.

\subsubsection{Example: $A(q,p)=q$.} 
\label{example1}
In this case,
\numparts
\begin{eqnarray}
\hat{A}_{op} &=& \frac{\partial A}{\partial q}\frac{\partial }{\partial p},
\label{A(q)} \\
\hat{A}_{op}^2 
&=& \left(\frac{\partial A}{\partial q}\right)^2\frac{\partial^2}{\partial p^2} 
\; ,
\label{A(q)->Aop}
\end{eqnarray}
\endnumparts
and Eqs. (\ref{CM diffus_eqn a}), (\ref{CM diffus_eqn b}) take the form
\begin{equation}
\frac{\partial \rho'_{s}(q,p)}{\partial \tau}
=  \left(\frac{\partial A}{\partial q}\right)^2 
\frac{\partial^2 \rho'_{s}(q,p)}{\partial p^2} \; .
\label{diffus_eqn 1}
\end{equation}
This is just the standard {\it diffusion equation}, where $(\partial A/\partial q)^2$ plays the role of a diffusion coefficient, which may be $q$-dependent, but is $p$-independent.

The CM diffusion equation (\ref{diffus_eqn 1}) has a structure similar to the
QM result for the Wigner function of the final system state, Eq. (\ref{diff_eqn_WT}), 
if the operator $\left(\frac{\partial A}{\partial q}\right)^2 
\frac{\partial^2}{\partial p^2}$ in the former is replaced by 
$\left[\frac{1}{i\hbar}\Delta A\left(q, i\hbar\frac{\partial}{\partial p}\right)\right]^2$ in the latter.
We observe that the two coincide in the particular case $A(q,p)=q$, or if we make the approximation
$\Delta A(q,y) \approx A'(q)y$ in Eq. (\ref{diff_eqn_WT}).

As is well known, the effect of diffusion is a Gaussian convolution: the solution of the diffusion equation (\ref{diffus_eqn 1}) is
\begin{equation}
\rho'_{s}(q,p)
=\int_{-\infty}^{\infty} 
\frac{e^{-\frac{\tilde{p}^2}{4 \tau\left(\frac{\partial A}{\partial q}\right)^2}}}
{\sqrt{4 \pi \tau \left(\frac{\partial A}{\partial q}\right)^2}}
\; \rho_s (q, p-\tilde{p}) d\tilde{p} .
\label{sol_diffus_eqn 1}
\end{equation}
From this result it follows that 
\begin{equation}
\int_{-\infty}^{\infty}  \rho'_{s}(q,p) dp  = \int_{-\infty}^{\infty}  \rho_s(q,p) dp,
\label{q-marginal_unchanged}
\end{equation}
so that the $q$-marginal distribution is not altered; this is similar to what occurs in QM, as we can see in  Eq. (\ref{w_a_nu unchanged}).
On the other hand, {\it the $p$-marginal experiences diffusion}:
for larger and larger $\tau$, $\rho'_{s\pi}(q,p)$ becomes more and more {\it a function of $q$ only}.
An analogous behavior occurs, in the strong-coupling limit, in QM, 
as can be seen in Eqs. (\ref{QM lueders 1}) and (\ref{diff_eqn_WT}) and the remarks following these equations.
It is striking to see that we have a 
{\it CM analogue of the QM L\"uders rule}.
This will be even more evident in the example \ref{example2} discussed below.

Notice that, whereas the CM initial distribution $\rho_s(q,p)$ is always non-negative, but otherwise arbitrary, the QM initial Wigner function
$W_{\hat{\rho}_{s}}(q,p)$
may fail, in general, to be non-negative, while it must certainly satisfy Heisenberg's uncertainty relations.
In the case $A(q,p)=q$, for the same initial distribution the resulting evolution is the same in CM as in QM.

From the above discussion we see that the ``hallmark" of a {\it non-selective} projective measurement of position, i.e., $A(q,p)=q$, both in CM and in QM, 
is diffusion in $p$ space of the classical distribution in the former, and of the WT in the latter.

\subsubsection{Example.}
\label{example2}

The diffusion process found in the previous example can be seen even more explicitly when the variable in which the diffusion occurs has a finite domain.
This case will be treated in detail in  \ref{meas_CM_heisenberg} in the ``Heisenberg picture"; 
here we just indicate some of the results within the ``Schr\"odinger picture'' that we have developed in the present section.

In terms of the variables $\bar{q}$, $\bar{p}$ defined in Eq. (\ref{qbar,pbar}) so as to have the same dimensions, assume that the system observable is a function of the combination 
$\xi=(\bar{p}^2+\bar{q}^2)/2$ of Eq. (\ref{xi}), i.e., 
$A(\bar{q}, \bar{p})=A(\xi)$, Eq. (\ref{A(xi)}).
In this case we have
\numparts
\begin{eqnarray}
\hat{A}_{op} &=& \frac{d A}{d \xi}
\frac{\partial}{\partial \theta},
\label{A(xi)op} \\
\hat{A}_{op}^2 
&=& \left(\frac{d A}{d \xi}\right)^2
\frac{\partial^2}{\partial \theta^2} \; ,
\label{A2(xi)op}
\end{eqnarray}
\label{A(xi)op,A2(xi)op}
\endnumparts
and the diffusion Eq. (\ref{CM diffus_eqn a}) takes the form
\numparts
\begin{eqnarray}
\frac{\partial \rho'_{s} (\xi, \theta)}{\partial \tau}
= \left(\frac{d A}{d \xi}\right)^2
\frac{\partial^2 \rho'_{s}(\xi, \theta)}{\partial \theta^2},
\label{drho/dtau}
\end{eqnarray}
to be solved with the ``initial condition"
\begin{eqnarray}
\rho'_{s,\tau=0}(\xi, \theta)=\rho_{s}(\xi, \theta).
\label{init.conds.}
\end{eqnarray}
\label{drho/dtau + init.conds.}
\endnumparts
\ref{A(ksi) CM} shows that the solution is
\begin{equation}
\rho'_{s}(\xi, \theta)
= \frac{1}{2 \pi} \sum_{m=-\infty}^{\infty} e^{im\theta}
e^{- m^2 \left(\frac{d A}{d \xi}\right)^2 \tau}
\int_0^{2\pi} e^{-im\theta'}
\rho_s(\xi, \theta') d\theta' .
\label{rhoA(xi,theta) 1}
\end{equation}
In the limit of very large coupling, only the term $m=0$ survives in the sum, and  we find
\numparts
\begin{eqnarray}
\lim_{\tau \to \infty}
\rho'_{s}(\xi, \theta) 
&=& 
\frac{1}{2 \pi} \int_0^{2\pi} \rho_s(\xi, \theta') d \theta'
= \frac{1}{2 \pi} \rho_s(\xi)  
\label{rhos tau to infty a}  \\
&=& \frac{1}{2 \pi} \int \int \delta(\xi - \xi')\rho_s(\xi', \theta') d\xi'd\theta' \; .
\label{rhos tau to infty b}
\end{eqnarray}
\label{rhos tau to infty}
\endnumparts

The $\theta$ marginal experiences diffusion.
Eq. (\ref{rhos tau to infty a}) shows that for larger and larger 
$\tau$, $\rho'_{s\pi}(\xi, \theta)$ becomes more and more isotropic in $\theta$ and, eventually,
{\it a function of $\xi$ only}.
Once again, this classical behavior reminds us of the QM one, expressed in 
Eqs. (\ref{QM lueders 1}) and (\ref{diff_eqn_WT}).
Eq. (\ref{rhos tau to infty b}), in turn, has a structure similar to the QM Eq. (\ref{QM lueders}) 
(which gives L\"uders rule for {\it non-selective} projective measurements),
the delta function playing the role of the projectors.
An illustrative example of this behavior is given in
 \ref{example of final system reduced rho CM}, which presents a situation similar to the QM one given in  \ref{example of final system reduced rho QM}.

\subsection{Selective projective measurements in CM}
\label{selective proj meas in CM}

Integrating Eq. (\ref{rhot>t1 b}) over $P$ we find
\begin{equation}
\rho'_{s \pi}(q,p,Q) 
= \int dP \Big[ e^{\epsilon \hat{A}_{op}(q,p)P}  \rho_s(q,p) \Big]
\rho_{\pi}(Q - \epsilon A(q,p),P) \; .
\label{rho' q,p,P 1}
\end{equation}
Assuming, for simplicity, that in the original probe state $P$ and $Q$ are statistically independent
\begin{equation}
\rho_{\pi}(Q,P)=\rho_{\pi}(Q)\rho_{\pi}(P),
\label{rhoM si}
\end{equation}
we have
\begin{equation}
\rho'_{s \pi}(q,p,Q) 
= \left\{ \int dP \Big[ e^{\epsilon \hat{A}_{op}(q,p)P} \rho_{\pi}(P) \Big] \rho_s(q,p) \right\}
\rho_{\pi}(Q - \epsilon A(q,p)) \; .
\label{rho' q,p,P 2}
\end{equation}
For a Gaussian distribution for $\rho_{\pi}(P)$, Eq. (\ref{rho'_s 0}) allows writing 
\begin{equation}
\rho'_{s \pi}(q,p,Q) 
= \left[ e^{\tau \hat{A}_{op}^2} \; \rho_s(q,p) \right]
\rho_{\pi}(Q - \epsilon A(q,p)) \; .
\label{rho' q,p,P gauss}
\end{equation}

In what follows we shall be interested in the conditional probability density of $q$ and $p$, on the condition that the probe coordinate is $Q$; 
it is given by
\begin{equation}
\rho'_{s}(q,p|Q)
=\frac{\left[ e^{\tau \hat{A}_{op}^2} \; \rho_s(q,p) \right]
\rho_{\pi}(Q - \epsilon A(q,p))}{\rho'_{\pi}(Q)} \; ,
\label{rho'(q,p|P) 1}
\end{equation}
where
\numparts
\begin{eqnarray}
\rho'_{\pi}(Q)
&=& \int \int dq dp
\left[ e^{\tau \hat{A}_{op}^2} \; \rho_s(q,p) \right]
\rho_{\pi}(Q - \epsilon A(q,p)) 
\label{rho'(Q) 1 a}   \\
&=& \int \int dq dp \;
\rho_s(q,p) \rho_{\pi}(Q - \epsilon A(q,p)) \; ,
\label{rho'(Q) 1 b}
\end{eqnarray}
\label{rho'(Q) 1}
\endnumparts
just as in Eq. (\ref{rho'(Q)}).
From Eq. (\ref{rho'(Q) 1 a}) to (\ref{rho'(Q) 1 b}) we have used an argument like the
one used in connection with Eqs. (\ref{rho(Q,P) 2})-(\ref{rhoM}).
As a simplification, assume $A(q,p)=q$. 
Then
\begin{equation}
\rho'_{s}(q,p|Q)
=\frac{\left[ e^{\tau \frac{\partial^2}{\partial p ^2}} \; \rho_s(q,p) \right]
\rho_{\pi}(Q - \epsilon q)}
{\int dq' \;\rho_s(q') \rho_{\pi}(Q - \epsilon q')} \; .
\label{rho'(q,p|P) 3}
\end{equation}
Suppose the width $\sigma_Q$ of $\rho_{\pi}(Q)$ is very small compared with intervals over which $\rho_s(q,p)$ varies appreciably with the variable $q$.
Then, for a given $Q$, $\rho_{\pi}(Q - \epsilon q)$ and so also the RHS of (\ref{rho'(q,p|P) 3}), considered as a function of $q$, is non-negligible only when 
$q \approx Q/\epsilon \pm \sigma_Q/\epsilon$;
in other words, for a given $Q$, $q \approx Q/\epsilon$ is {\it selected}.
Thus, choosing for $Q$ some fixed value,
i) in the first factor in the numerator of Eq. (\ref{rho'(q,p|P) 3}) we can replace 
$q\approx Q/\epsilon$, since, away from $q \approx Q/\epsilon \pm \sigma_Q/\epsilon$,
the second factor, i.e., $\rho_{\pi}(Q - \epsilon q)$, is negligible anyway; 
ii) in the denominator of Eq. (\ref{rho'(q,p|P) 3}), $\rho_s(q')$ varies much more slowly than 
$\rho_{\pi}(Q - \epsilon q')$; we can thus replace $q'\approx Q/\epsilon$ in $\rho_s(q')$ and take it out of the integral.

We thus obtain
\begin{equation}
\rho'_s(q,p|Q)
\approx \frac{\left[ e^{\tau \frac{\partial^2}{\partial p ^2}} \; 
\rho_s(Q/\epsilon,p) \right]
\rho_{\pi}(Q - \epsilon q)}
{\rho_s(Q/\epsilon) \int  dq' \; \rho_{\pi}(Q - \epsilon q')} \; .
\end{equation}
Choosing $Q=\epsilon q_0$,
\numparts
\begin{eqnarray}
\rho'_s(q,p|Q=\epsilon q_0)
&\approx& \frac{\left[ e^{\tau \frac{\partial^2}{\partial p ^2}} \; \rho_s(q_0,p) \right]
\rho_{\pi}(q - q_0)}
{\rho_s(q_0)} \\
&\approx& \frac{ e^{\tau \frac{\partial^2}{\partial p ^2}} \; \rho_s(q_0,p)}
{\rho_s(q_0)}
\delta(q - q_0) \; .
\label{rho'(q,p|P) 4}
\end{eqnarray}
\endnumparts
Notice the similarity of this result with that of Eq.~(\ref{rho's|Q 2}), which describes the QM L\"uders rule for a {\it selective projective measurement}, sometimes referred to as ``collapse".
Indeed, for large values of $\tau$ the result~(\ref{rho'(q,p|P) 4}) 
has a definite position $q_0$ and has suffered an appreciable diffusion in momentum, in a similar way as to what occurs in QM.
This is exhibited in the two marginals that can be computed from Eq. (\ref{rho'(q,p|P) 4}):
\begin{equation}
\int dp \rho'_s(q,p|Q=\epsilon q_0)
\approx 
\frac{\int dp \; \rho_s(q_0,p)} {\rho_s(q_0)}
\delta(q - q_0)
= \delta(q - q_0) 
\end{equation}
(where only the first term in the expansion of the exponential survives the integration over $p$), and 
\begin{equation}
\int dq \rho'_s(q,p|Q=\epsilon q_0)
\approx 
\frac{e^{\tau \frac{\partial^2}{\partial p ^2}} \; \rho_s(q_0,p)}
{\rho_s(q_0)}
= e^{\tau \frac{\partial^2}{\partial p ^2}} \rho_s(p|q_0) \; .
\end{equation}

\section{Summary and Conclusions}
\label{conclusions}

In this paper we considered non-ideal measurements on classical systems. 
In particular, we assumed that the system of interest is coupled to a probe whose initial phase-space distribution is different from a Dirac delta function, so as to represent a statistical uncertainty.
The probe describes an early stage of the measurement, which might be called a ``pre-measurement" and is described dynamically in our model.
The probe is eventually detected to complete the measurement.

We found that there is a great similarity in structure between the QM and the CM problems, provided that, in the latter, we perform non-ideal measurements. 
We summarize this similarity in Table~\ref{tbl:sum}. 

\begin{table}[ht]
\begin{tabular}{| p{3.4cm}| l | l |}
\hline
&Quantum mechanics & Classical mechanics\\
\hline\hline
Expectation values&Eq.~(\ref{<Q>QM}): $\frac{\langle Q \rangle'}{\epsilon} 
= \langle A  \rangle$& Eq.~(\ref{<Q>'CM}): $\frac{\langle Q \rangle'}{\epsilon} 
= \langle A  \rangle$\\\hline
Uncertainty&Eq.~(\ref{QM-uncert-x}): $\sim \frac{\sigma_Q}{\epsilon}$ & Eq.~(\ref{CM-uncert-q}): $\sim \frac{\sigma_Q}{\epsilon}$\\\hline
Final reduced state& Eq.~(\ref{QM rho's 1}): $\hat{\rho}'_{s} = e^{- \frac{\tau}{\hbar^2} \hat{\cal{A}}^2} \hat{\rho}_{s}$ & Eq.~(\ref{rho'_s 1}): $\rho'_{s}(q,p)
=  e^{\tau \hat{A}_{op}^2} \; \rho_s(q,p)$ \\\hline
Diffusion equation& Eq.~(\ref{lindblad}): $\frac{\partial \hat{\rho}'_{s}}{\partial \tau} 
= - \frac{1}{\hbar^2} \hat{\cal{A}}^2 \hat{\rho}'_{s}$ & 
Eq.~(\ref{CM diffus_eqn a}): $\frac{\partial \rho'_{s}(q,p)}{\partial \tau}
= \hat{A}_{op}^2 \rho'_{s}(q,p)$ \\
\hline\hline
\end{tabular}
\caption{Structural similarities between measurements in QM and CM.}
\label{tbl:sum}
\end{table}%

The structural similarity of the equations describing the system-probe evolution in QM and in CM imply that several aspects of the physical behavior are similar.
A striking example is the reduced state $\rho'_{s}$ of the system proper after its interaction with the probe.
As a function of the coupling strength $\tau$, in QM it ``evolves" according to the Lindblad equation; in CM it evolves, basically, according to a diffusion equation.
As a result, for a given system observable $\hat{A}$ the distribution of the canonically conjugate variable experiences diffusion as a consequence of its interaction with the probe, which is seen in the Wigner function in the QM case, and in the classical distribution in phase space in the CM one. 
In the strong-coupling limit, $\tau \to \infty$, one finds, in QM, L\"uders rule for a {\it non-selective} projective measurement, where the resulting density operator is a function of the observable $\hat{A}$ alone;
a similar result can be seen in the limit of the CM $\rho'_{s}$.

In a variant of the model, one can also describe a {\it selective} projective measurement which, in QM, gives rise to the so-called {\it collapse} of the state.
In CM a very similar result is encountered, where one finds a definite value of the selected dynamical variable, while the canonically conjugate one has suffered diffusion.

Based on our analysis, we conclude that a number of features of the measurement process which are sometimes considered as unique to quantum theory, may be seen as phenomena that are directly related to the statistical nature of the theory, and indeed can be found in a probabilistic classical theory as well.

\ack
AK acknowledges support from NSF Grant No. PHY-$1212445$, and 
PAM from Conacyt Grant No. 79501.
PAM is also grateful to the Faculty of Technology of the Buskerud University College, Kongsberg, Norway, and to the
Centre for Quantum Technologies, National University of Singapore, Singapore, where part of this work was carried out. 

\appendix

\section{Example of Eq. (\ref{QM lueders 1})}
\label{example of final system reduced rho QM}

Suppose the system operator of interest is taken as
\begin{eqnarray}
\hat{A}&=& \hat{\xi} 
= \frac12 (\hat{\bar{p}}^2 + \hat{\bar{q}}^2) 
\label{AQM=xi a} \\
&=& \left(\hat{a}^{\dagger}\hat{a} + \frac12\right) \hbar 
= \left(\hat{n} + \frac12\right)\hbar ,
\label{AQM=xi b} 
\end{eqnarray}
where
\begin{eqnarray}
\hat{a} 
&=& \frac{\hat{\bar{q}} + i \hat{\bar{p}}}{\sqrt{2 \hbar}},
\hspace{1cm}
\hat{a}^{\dagger} 
= \frac{\hat{\bar{q}} - i \hat{\bar{p}}}{\sqrt{2 \hbar}}
\label{a+,a,n} \\
\hat{n} &=& \hat{a}^{\dagger} \hat{a} , 
\hspace{1cm}
\hat{n} |n \rangle = n |n \rangle,
\label{n}
\end{eqnarray}
The operators $\hat{\bar{q}}$ and $\hat{\bar{p}}$ are the QM version of the classical variables of Eq. (\ref{qbar,pbar}).
The operator $\hat{\xi}$ has the spectral representation
\begin{equation}
\hat{\xi} 
= \hbar \sum_n |n \rangle \left(n+\frac12\right) \langle n| .
\label{xi, QM}
\end{equation}
The state of the system at $t=0$ is taken as 
\begin{eqnarray}
\hat{\rho}_s
&=& 2 \sinh \left(\frac{\hbar}{2\sigma_{\bar{p}} \sigma_{\bar{q}}} \right)
e^{-\frac12 \left(\frac{\hat{\bar{p}}^2}{\sigma_{\bar{p}}^2} + \frac{\hat{\bar{q}}^2}{\sigma_{\bar{q}}^2} \right)}   
\label{rho0s qm 1 a} \\
&=& 2 \sinh \left(\frac{\hbar}{2\sigma_{\bar{p}} \sigma_{\bar{q}}} \right)
e^{-\frac{1}{\sigma_{\bar{p}}^2} 
\left[
\hat{\xi}  
+ \frac12 \left(\frac{\sigma_{\bar{p}}^2}{\sigma_{\bar{q}}^2}-1 \right) \hat{\bar{q}}^2
\right]} \; .
\label{rho0s qm 1 b}
\end{eqnarray}

From Eq. (\ref{QM lueders 1}), the reduced density operator for the system proper becomes, in the strong-coupling limit 
\begin{equation}
\lim_{\epsilon \sigma_P \to \infty} \hat{\rho}'_{s}
= \sum_{n} |n \rangle p_{s}(n)\langle n | ,
\label{QM_Lueder 2}
\end{equation}
with
\begin{equation}
 p_{s}(n)
= 2 \sinh \left(\frac{\hbar}{2\sigma_{\bar{p}} \sigma_{\bar{q}}} \right)
\left\langle n \left|
e^{-\frac{1}{\sigma_p^2} 
\left[
\hat{\xi}  
+ \frac12 \left(\frac{\sigma_{\bar{p}}^2}{\sigma_{\bar{q}}^2}-1 \right) \hat{\bar{q}}^2
\right]}
\right| n \right\rangle \; .
\label{QM_Lueder 3}
\end{equation}

While the original system density operator $\hat{\rho}_s$ of Eq. (\ref{rho0s qm 1 b}) is not a function of $\hat{\xi}$ only, the final one after the interaction with the probe is over, Eq. (\ref{QM_Lueder 2}),
{\it is} only a function of $\hat{\xi}$ in the strong-coupling limit,
as is seen comparing its spectral representation with that of $\hat{\xi}$, 
Eq. (\ref{xi, QM}).

\section{Proof of Eq.~(\ref{[AP,AdQ]}).}
\label{proof_[AP,AdQ]}

From its definition, the operator $\hat{A}_{op}(q,p)$ has the property
\begin{equation}
\hat{A}_{op}(q,p) A(q,p) 
\equiv [A(q,p), A(q,p)]_{PB} =0.
\label{[A,A]=0}
\end{equation}
For any two functions $f(q,p)$, $g(q,p)$, the operator $\hat{A}_{op}(q,p)$ also
has the property:
\begin{eqnarray}\label{distr_Aop}
\hat{A}_{op} (f g)
&\equiv& [A, fg]_{PB} \nonumber\\
&=& [A, f]_{PB} \; g + f \; [A, g]_{PB}\nonumber\\
&=& (\hat{A}_{op} f) g   +  f(\hat{A}_{op} g) .
\end{eqnarray}
Assuming an arbitrary function $F(q,p,Q,P)$, we can write, on the one hand
\begin{eqnarray}
\left[(\hat{A}_{op}(q,p) P) (A(q,p) \partial_Q)\right] F(q,p,Q,P)
&=& P \partial_Q \hat{A}_{op}(AF) 
\label{APAdQ a}\\
&=&  P \partial_Q A \hat{A}_{op} F
\label{APAdQ b} \\
&=& (PA \hat{A}_{op} \partial_Q) F .
\label{APAdQ c}
\end{eqnarray}
From Eq. (\ref{APAdQ a}) to Eq. (\ref{APAdQ b}) we have used properties 
(\ref{distr_Aop}) and (\ref{[A,A]=0}).
On the other hand, we can also write
\begin{equation}
\left[(A(q,p) \partial_Q) (\hat{A}_{op}(q,p) P)\right]F(q,p,Q,P)
= (AP \hat{A}_{op} \partial_Q) F .
\label{AdQAP}
\end{equation}
From Eqs.~(\ref{APAdQ a})-(\ref{AdQAP}), the desired result (\ref{[AP,AdQ]})
follows.


\section{Derivation of Eq. (\ref{rhoA(xi,theta) 1})}
\label{A(ksi) CM}

If we write, for the solution of (\ref{drho/dtau}) and (\ref{init.conds.}), the expansion
\begin{eqnarray}
\rho'_{s}(\xi, \theta)
= \sum_m c_m(\tau, \xi) e^{im\theta},
\label{expansion of rhoA(tau)}
\end{eqnarray}
we find, for the expansion coefficients, the differential equation
\begin{eqnarray}
\frac{\partial c_m(\tau, \xi)}{\partial \tau}
= - \left(\frac{dA}{d\xi}\right)^2 m^2 c_m(\tau, \xi),
\label{dcm/dtau}
\end{eqnarray}
with the solution
\begin{eqnarray}
c_m(\tau, \xi)
= c_m(\tau=0, \xi) \; e^{-m^2 \left(\frac{dA}{d\xi}\right)^2 \tau} .
\label{cm(tau)}
\end{eqnarray}
The coefficients $c_m(\tau=0, \xi)$ can be obtained from the expansion of the initial probability density $\rho_{s}(\xi, \theta)$, i.e.
\begin{eqnarray}
\rho_{s} (\xi, \theta)
&=& \sum_m c_m(\tau=0, \xi) e^{im\theta}  \\
c_m(\tau=0, \xi) 
&=& \frac{1}{2 \pi}
\int_0^{2 \pi}\rho_{s} (\xi, \theta) e^{-i m \theta} d \theta.
\label{cm(0)}
\end{eqnarray}
We substitute the coefficient $c_m(\tau=0, \xi)$ of (\ref{cm(0)}) in (\ref{cm(tau)}), and the result in (\ref{expansion of rhoA(tau)}), to obtain Eq. (\ref{rhoA(xi,theta) 1}).

\section{Example of Eq. (\ref{rhos tau to infty a})}
\label{example of final system reduced rho CM}

As an example of the construction of the final distribution 
(\ref{rhos tau to infty a}) in the strong-coupling limit, suppose that the system proper has, originally, the Gaussian distribution (in the notation of Eq. (\ref{qbar,pbar}))
\begin{equation}
\rho_{s}(\bar{q},\bar{p})d\bar{p} d\bar{q}
= \frac{e^{-\frac12\left(\frac{\bar{p}^2}{\sigma_{\bar{p}}^2} + \frac{\bar{q}^2}{\sigma_{\bar{q}}^2} \right)}}{2 \pi \sigma_{\bar{p}} \sigma_{\bar{q}}} d\bar{p} d\bar{q}.
\label{rho0s 1}
\end{equation}
From Eqs. (\ref{xi}) and (\ref{theta}) we find the relations
\begin{eqnarray}
\bar{q} = \sqrt{2 \xi} \cos \theta \\
\bar{p} = \sqrt{2 \xi} \sin \theta,
\end{eqnarray}
so that the original distribution of the canonically related variables $\xi, \theta$ is
\begin{equation}
\rho_{s}(\xi,\theta)d\xi d\theta
= \frac{e^{- \frac{\xi}{\sigma_{\bar{p}}^2}
\left[  
1+\left(\frac{\sigma_{\bar{p}}^2}{\sigma_{\bar{q}}^2}-1\right) \cos^2 \theta   
\right]}}{2 \pi \sigma_{\bar{p}} \sigma_{\bar{q}}} d\xi d\theta,
\label{rho0s 2}
\end{equation}
with its $\xi$-marginal distribution given by
\begin{eqnarray}
\rho_{s}(\xi)
&=& \int_0^{2 \pi} \rho_{s}(\xi,\theta)d\xi d\theta \\
&=&\frac{e^{-\frac{\xi}{2\sigma_{\bar{p}}^2}\left(\frac{\sigma_{\bar{p}}^2}{\sigma_{\bar{q}}^2}+1\right)}}
{\sigma_{\bar{p}} \sigma_{\bar{q}}}
I_0\left(\frac{\xi}{2\sigma_{\bar{p}}^2}\left(\frac{\sigma_{\bar{p}}^2}{\sigma_{\bar{q}}^2}-1\right)\right) \; ,
\end{eqnarray}
$I_0$ being a modified Bessel function \cite{abramowitz}.
From Eq. (\ref{rhos tau to infty a}) we find, in the limit of {\it strong coupling}, the $\theta$-independent final distribution of the system proper as
\begin{equation}
\rho'_{s}(\xi, \theta)
=\frac{e^{-\frac{\xi}{2\sigma_{\bar{p}}^2}\left(\frac{\sigma_{\bar{p}}^2}{\sigma_{\bar{q}}^2}+1\right)}}
{2\pi \sigma_{\bar{p}} \sigma_{\bar{q}}}
I_0\left(\frac{\xi}{2\sigma_{\bar{p}}^2}\left(\frac{\sigma_{\bar{p}}^2}{\sigma_{\bar{q}}^2}-1\right)\right) \; .
\label{rhoA}
\end{equation}

\section{Measurements in Classical Mechanics:
the Heisenberg picture}
\label{meas_CM_heisenberg}
The problem of a classical measurement is considered in Sec. \ref{sec:cm} in what we might call the ``Schr\"odinger picture": the state, defined as the density in the phase-space of the system plus the probe, changes with time in accordance with Liouville's Eq. (\ref{Liouv_eqn}), while the observables stay fixed.
The results are quite general, i.e., applicable to arbitrary observables.

However, when the observables are sufficiently simple, we may gain an intuitive insight by considering the ``Heisenberg picture", in which the observables evolve in time according to Hamilton's equations.
Here we illustrate this fact in two particularly transparent cases.

We assume that the model Hamiltonian is given by a slight generalization of  Eq. (\ref{V_vNM CM}), 
\begin{equation}
H(t) = \epsilon  g(t)A(q,p) P \; ,
\label{V_vNM CM 1}
\end{equation}
where $\delta(t-t_1)$ has been replaced by the function $g(t)$, which is normalized to 1 and has a compact support around the interaction time $t_1>0$, i.e.,
\begin{equation}
G(t) = \int_0^t g(t') dt' ,  \;\;\; {\rm with} \;\;\; 
G(t \gg t_1) =1.
\label{G(t)}
\end{equation}
We analyze two particular examples.

\noindent\emph{Example 1: $A(q,p)=q$.}

Consider the observable $A(q,p)=q$.
Hamilton's equations give the equations of motion
\begin{eqnarray}
\dot{q} &=& 
0, \hspace{20mm}
\dot{p} 
= -\epsilon g(t) P ,
\label{dot-p} \\
\dot{Q} &=& 
\epsilon g(t) q  ,
\hspace{1cm}
\dot{P} 
=0 .
\label{dot-P 0}
\end{eqnarray}
We see that $q(t)$ and $P(t)$ are constants of the motion, i.e.,
\begin{eqnarray}
q(t) &=& q_0,  \;\;\;\;\;\;\; q' \equiv q(t \gg t_1) = q_0
\label{q(t)} \\
P(t) &=& P_0, \;\;\;\;\; P' \equiv P(t \gg t_1) = P_0 \; ,
\label{P(t) 0}
\end{eqnarray}
whereas $p(t)$ and $Q(t)$ vary in time as
\begin{eqnarray}
p(t) &=& p_0 - \epsilon P_0 G(t) ; \;\;\;\;\;\;
p' = p_0 - \epsilon P_0 ,
\label{p(t)} \\
Q(t) &=& Q_0 + \epsilon q_0 G(t) ; \;\;\;\;\;
Q' =Q_0 + \epsilon q_0 \; .
\label{Q(t) 0}
\end{eqnarray}
As in the text, the ``prime" indicates that the variable is evaluated at a time where the interaction $g(t)$ has ceased to act.

Thus, the system position $q$ is unchanged, while the change in the system momentum $p$ is
$\;-\epsilon P_0$, which vanishes when $P_0=0$:
in this case both the position and momentum distributions of the system proper are undisturbed by the measurement.
This is consistent with the comment made right below 
Eq. (\ref{rho'_s 1}), that the marginal distribution of the system remains unchanged by the measurement when $\langle P\rangle = \sigma_P=0$, 
i.e., when the probe has zero momentum.
If $P_0$ has an RMS value $\sigma_P$ around $P_0=0$, we can estimate the
\begin{equation}
\{ {\rm disturbance \; caused \; in} \; p \}
\sim \epsilon \; \sigma_P \; ,
\label{disturbance in p}
\end{equation}
a result reminiscent of the QM one, Eq. (\ref{QM disturbance on system}).
Using Eqs. (\ref{CM-uncert-q}) and (\ref{disturbance in p}), we can write
\begin{eqnarray}
&&\{{\rm uncertainty \; in \; discriminating} \; q^{,}{\rm s} 
{\rm \; in \; the \; original \; distribution} \} 
\nonumber \\
&& \hspace{5cm} \times \; \{ {\rm disturbance \; caused\; in \;} p \} 
\sim \sigma_Q \sigma_P \; .
\label{dq.dp}
\end{eqnarray}
Of course, in the CM case the RHS of this last result can be arbitrary, so that, 
if we make the uncertainty in discriminating $q^{,}$s very small, we are not compelled to produce a large disturbance in the momentum of the system proper,
in contrast to what occurs in QM.

We assume that at $t=0$ the statistical distribution of the original dynamical variables 
$q_0, p_0, Q_0, P_0$ is 
$\rho_{s}(q_0,p_0) \rho_{\pi}(Q_0,P_0)$, just as in Sect. \ref{sec:cm}.
The final probe position $Q'$, given in Eq. (\ref{Q(t) 0}), is then distributed as
\begin{eqnarray}
\rho'_{\pi}(Q')
&=&\int\int\int\int \delta(Q'-(Q_0+\epsilon q_0))
\rho_{s}(q_0,p_0) \rho_{\pi}(Q_0,P_0) dq_0 dp_0 dQ_0 dP_0 \nonumber\\
&=&\int \rho_{s}(q_0)\rho_{\pi}(Q'-\epsilon q_0) dq_0 \; ,
\end{eqnarray}
just as in Eq. (\ref{rho(Q) A=q}) obtained in the Schr\"odinger picture, replacing the present $Q'$ and $q_0$ by $Q$ and $q$, respectively.

Similarly, the final system variables $q',p'$ are distributed as
\begin{eqnarray}
&&\rho'_s(q',p') \nonumber\\&=& \int\int\int\int \delta(q'-q_0)  \delta(p'-(p_0-\epsilon P_0))
\rho_{s}(q_0,p_0) \rho_{\pi}(Q_0,P_0) dq_0 dp_0 dQ_0 dP_0
\nonumber \\ 
&=& \int \rho_{\pi}(P_0)\rho_{s}(q',p'+\epsilon P_0) dP_0 .
\label{rho'(q',p') heisenberg a,b}
\end{eqnarray}
This result coincides with Eq. (\ref{rho_s_t>t1 1}) obtained in the Schr\"odinger picture, in the particular case $A(q,p)=q$, replacing the present $q',p'$ by $q,p$.
Assuming for $P_0$ a Gaussian distribution, as in Eq. (\ref{rho(P)}), we have
\begin{equation}\label{rho'(q',p') heisenberg c} 
\rho'_s(q',p')
=\int \frac{{\rm e}^{-\frac{P_0^2}{2 \sigma_{P_0}^2}}}{\sqrt{2\pi\sigma_{P_0}^2 }}
\rho_{s}(q',p'+\epsilon P_0) dP_0 
= \int_{-\infty}^{\infty} 
\frac{{\rm e}^{-\frac{\bar{p}^2}{4 \tau}}}{\sqrt{4\pi\tau}}
\rho_{s}(q',p'- \tilde{p}) d\tilde{p}, \;\;
\end{equation}
just as in Eq. (\ref{sol_diffus_eqn 1}) for $A(q,p)=q$.

\noindent\emph{Example 2: 
$A(\bar{q}, \bar{p})= A\left(\frac12(\bar{p}^2+\bar{q}^2)\right)$.}

For convenience, we first define the units of our dynamical variables as follows. 
Starting from $q$ and $p$, we make a canonical transformation
\begin{equation}
\bar{q}=C q, \hspace{5mm} \bar{p}=\frac{p}{C} \; ,
\label{qbar,pbar}
\end{equation}
so that $\bar{q}$ and $\bar{p}$ have the same dimensions: then $C$ has dimensions of 
$[C]=[\sqrt{h}/q]$ and $[\bar{q}]=[\bar{p}]=[\sqrt{h}]$.
A similar procedure defines $\bar{Q}$ and $\bar{P}$ with dimensions of $\sqrt{h}$.

In the present example, we choose the observable $A(\bar{q},\bar{p})$ to be a function of the combination $\xi$ defined as
\begin{eqnarray} 
\xi &=& \frac12(\bar{p}^2+\bar{q}^2) \; ,
\label{xi}
\end{eqnarray}
so that
\begin{eqnarray}
A(\bar{q},\bar{p}) &=& A(\xi) \; .
\label{A(xi)}
\end{eqnarray}
The variable $\xi$ is canonically conjugate to the variable $\theta$, i.e.,
\begin{eqnarray}
&&\theta = \tan^{-1}\frac{\bar{p}}{\bar{q}} \; ,
\label{theta} \\
&&\left[\xi, \theta \right]_{PB} = 1 .
\label{[xi,theta]}
\end{eqnarray}
The Hamiltonian will be taken to be
\begin{eqnarray}
H(t) = \epsilon g(t) A(\xi)\bar{P}.
\label{H(ksi,P,t)} 
\end{eqnarray}
The particular feature of the present example is that the variable $\theta$ has a {\it finite domain}, so that diffusion in this variable will eventually produce an isotropic distribution.

From Hamilton's equations we find the equations of motion
\begin{eqnarray}
\dot{\xi} &=& 0 , 
\hspace{30mm}
\dot{\theta} 
= -\epsilon g(t) \frac{\partial A}{\partial \xi} \bar{P} ,
\label{dot-xi,theta} \\
\dot{\bar{Q}} &=& 
= \epsilon g(t) A(\xi) ,
\hspace{1cm}
\dot{\bar{P}} = 0 .
\label{dot-Q,P}
\end{eqnarray}
We see that $\xi(t)$ and $\bar{P}(t)$ are constants of the motion, i.e.,
\begin{eqnarray}
\xi(t) &=& \xi_0,  \;\;\;\;\;\;\; \xi' \equiv \xi(t \gg t_1) = \xi_0
\label{xi(t)} \\
\bar{P}(t) &=& \bar{P}_0, \;\;\;\;\; \bar{P}' \equiv \bar{P}(t \gg t_1) 
= \bar{P}_0 \; ,
\label{P(t)}
\end{eqnarray}
whereas $\theta(t)$ and $Q(t)$ vary in time as
\begin{eqnarray}
\theta(t) &=& \theta_0 - \epsilon 
\left(\frac{\partial A}{\partial \xi}\right)_{\xi_0} P_0 G(t) ; \hspace{12mm}
 \theta' = \theta_0 - 
 \epsilon \left(\frac{\partial A}{\partial \xi}\right)_{\xi_0} P_0 ,
\label{theta(t)} \\
\bar{Q}(t) &=& \bar{Q}_0 + \epsilon A(\xi_0) G(t) ; \hspace{21mm}
\bar{Q}' =\bar{Q}_0 + \epsilon A(\xi_0) \; .
\label{Q(t)}
\end{eqnarray}
Here, as before, the ``prime" indicates that the variable is evaluated at a time where the interaction $g(t)$ has ceased to act.

We assume that at $t=0$ the statistical distribution of the original dynamical variables
$\bar{q}_0, \bar{p}_0, \bar{Q}_0, \bar{P}_0$ is
$\rho_{s}(\bar{q}_0, \bar{p}_0) \rho_{\pi}(\bar{Q}_0, \bar{P}_0)$, just as in Sect. \ref{sec:cm}.
Thus the final probe position $\bar{Q}'$, given in Eq. (\ref{Q(t)}), is distributed as
\begin{eqnarray}
&&\rho'_{\pi}(\bar{Q}') \nonumber\\
&=& \int\int\int\int
\delta(\bar{Q}'-(\bar{Q}_0 + \epsilon A(\bar{q}_0,\bar{p}_0))) 
\rho_{s}(\bar{q}_0,\bar{p}_0) \rho_{\pi} (\bar{Q}_0,\bar{P}_0) 
d\bar{q}_0 d\bar{p}_0 d\bar{Q}_0 d\bar{P}_0
\label{p(Q) 2a} 
\nonumber \\ 
&=& \int \int \rho_{s}(\bar{q}_0,\bar{p}_0) \rho_{\pi}
\left(\bar{Q}'-\epsilon A\left(\frac{\bar{q}_0^2+\bar{p}_0^2}{2}\right)\right) d\bar{q}_0 d\bar{p}_0.
\label{p(Q) 2}
\end{eqnarray}
Using the fact that $\bar{q}_0,\bar{p}_0$ and $\xi_0, \theta_0$ are related by a canonical transformation, (\ref{p(Q) 2}) can also be expressed as
\begin{eqnarray}
\rho'_{\pi}(\bar{Q}') =  
\int \int \rho_{s}(\xi_0, \theta_0) \rho_{\pi}(\bar{Q}'-\epsilon A(\xi_0)) 
d\xi_0 d\theta_0.
\label{p(Q) 3}
\end{eqnarray}
Eqs. (\ref{p(Q) 2}) and (\ref{p(Q) 3}) are consistent with the result (\ref{rho'(Q)}) obtained in the Schr\"odinger picture.
The integration over $\theta_0$ can be performed, giving the marginal distribution of $\xi_0$ at $t=0$, $\rho_{s}(\xi_0)$, so that
\begin{eqnarray}
\rho'_{\pi}(\bar{Q}') =  
\int \rho_{s}(\xi_0) \rho_{\pi}(\bar{Q}'-\epsilon A(\xi_0)) d\xi_0 .
\label{p(Q) 4}
\end{eqnarray}
The expectation value of the probe position after the interaction is over is then given by
\begin{equation}
\frac{\langle \bar{Q}' \rangle}{\epsilon} 
= \int A(\xi_0) \rho_{s}(\xi_0) d\xi_0
=  \langle  A(\xi_0) \rangle \; ,
\label{<Q>} 
\end{equation}
(where we have assumed that the original expectation value of $\bar{Q}_0$ vanishes),
a result to be compared with the general one, Eq. (\ref{<Q>'CM}), obtained in the Schr\"odinger picture.


We now concentrate on the final marginal distribution of the system proper. 
Eqs. (\ref{xi(t)}) show that the system variable $\xi$ is unchanged by the interaction, so that its marginal distribution remains unaltered by the measurement process, i.e.,
\begin{equation}
\rho'_{s}(\xi) = \rho_{s}(\xi).
\label{p(xi) 1}
\end{equation}
On the other hand, Eq. (\ref{theta(t)}) shows that the system variable $\theta$ suffers a change due to the interaction with the probe.
From Eqs. (\ref{xi(t)}) and (\ref{theta(t)}) we find the final marginal joint probability density of $\xi$ and $\theta$ as
\begin{eqnarray}
&&\rho'_{s} (\xi', \theta')\nonumber\\
 &=&\int \int \int \int
\delta({\xi' - \xi_0}) 
\delta\left(\theta' - \left(\theta_0 - \epsilon 
\left(\frac{\partial A}{\partial \xi}\right)_{\xi_0} \bar{P}_0 \right) \right) \nonumber\\
&&\times\rho_{s}(\xi_0, \theta_0) \rho_{\pi}(\bar{Q}_0,\bar{P}_0) d\xi_0 d\theta_0 d\bar{Q}_0 d\bar{P}_0  
\nonumber \\ 
&=& \frac{1}{2 \pi} \sum_{m=-\infty}^{\infty} e^{im\theta'}
\int \int e^{-im\left[\theta_0 - \epsilon 
\left(\frac{\partial A}{\partial \xi}\right)_{\xi'} \bar{P}_0\right]}
\rho_{s}(\xi', \theta_0) \rho_{\pi}(\bar{P}_0)d\theta_0 d\bar{P}_0 \;. 
\end{eqnarray}
For a Gaussian for the marginal distribution $\rho_{\pi} (\bar{P}_0)$ of $\bar{P}_0$ at $t=0$, we have
\begin{equation}
\int_{-\infty}^{\infty} e^{im\epsilon 
\left(\frac{\partial A}{\partial \xi}\right)_{\xi'} \bar{P}_0}\rho_{\pi}(\bar{P}_0) d\bar{P}_0
=e^{-\frac12 m^2 \epsilon^2 \sigma_{\bar{P}_0}^2
\left[\left(\frac{\partial A}{\partial \xi}\right)_{\xi'}\right]^2} ,
\label{<exp(imeAP_0)>P_0}
\end{equation}
so that
\begin{eqnarray}
\rho'_{s} (\xi', \theta')
= \frac{1}{2 \pi} \sum_{m=-\infty}^{\infty} e^{im\theta'}
e^{-\frac12 m^2 \epsilon^2 \sigma_{\bar{P}_0}^2 
\left[\left(\frac{\partial A}{\partial \xi}\right)_{\xi'}\right]^2}
\int_0^{2\pi} e^{-im\theta_0}
\rho_{s}(\xi', \theta_0) d\theta_0,
\label{rhoA(xi,theta) d}
\end{eqnarray}
just as in Eq. (\ref{rhoA(xi,theta) 1}), obtained in the Schr\"odinger picture.


\end{document}